\documentclass[twocolumn,showpacs,superscriptaddress,preprintnumbers]{revtex4-1}
\usepackage{amsfonts, amsmath, amssymb, bm, bbm, graphicx, epsfig, epstopdf}
\usepackage{gensymb}
\usepackage{dcolumn}
\usepackage{textcomp}
\usepackage{hyperref}
\usepackage{color}
\usepackage{natbib}
\usepackage[utf8]{inputenc}    
\usepackage[normalem]{ulem}
\usepackage{mathtools}
\usepackage{epstopdf}
\usepackage{amsmath}
\usepackage{csquotes}
\usepackage{hyperref}
\usepackage{url}

\def\opone{\leavevmode\hbox{\small1\kern-3.8pt\normalsize1}}

\begin{document}

\title{Practical quantum random number generator based on sampling vacuum fluctuations}

\author{Qiang Zhou}
\affiliation{Institute of Fundamental and Frontier Science, University of Electronic Science and Technology of China (UESTC), Chengdu 610054, China and the School of Optoelectronic Science and Engineering, UESTC, Chengdu 610054, China}
\affiliation{Institute for Quantum Science and Technology, and Department of Physics and Astronomy, University of Calgary, Calgary T2N 1N4, Alberta, Canada}

\author{Raju Valivarthi}
\affiliation{ICFO-Institut de Ciences Fotoniques, Castelldefels, E-08860 Barcelona, Spain}
\affiliation{Institute for Quantum Science and Technology, and Department of Physics and Astronomy, University of Calgary, Calgary T2N 1N4, Alberta, Canada}

\author{Caleb John}
\affiliation{Department of Electrical and Computer Engineering, University of Calgary, Calgary, AB, T2N 1N4, Canada}

\author{Wolfgang Tittel}
\affiliation{QuTech, Delft University of Technology, 2611 LC Delft, Netherlands}
\affiliation{Institute for Quantum Science and Technology, and Department of Physics and Astronomy, University of Calgary, Calgary T2N 1N4, Alberta, Canada}

\date{\today}

\maketitle
Random number generation is an enabling technology for fields as varied as Monte Carlo simulations and quantum information science.~An important application is a secure quantum key distribution (QKD) system; here, we propose and demonstrate an approach to random number generation that satisfies the specific requirements for QKD.~In our scheme, vacuum fluctuations of the electromagnetic-field inside a laser cavity are sampled in a discrete manner in time and amplified by injecting current pulses into the laser.~Random numbers can be obtained by interfering the laser pulses with another independent laser operating at the same frequency.~Using only off-the-shelf opto-electronics and fibre-optics components at 1.5~$\mu$m wavelength,~we  experimentally demonstrate the generation of high-quality random bits at a rate of up to 1.5 GHz.~Our results show the potential of the new scheme for practical information processing applications.

\section{Introduction}

The generation of true random numbers is highly desirable for digital information systems \cite{Bennett1984,Metropolis1949,Schneier1995}. For instance, in quantum key distribution (QKD), random bits are used as a seed for creating secure keys shared between two legitimate users \cite{Gisin2002,Gisin2007,Lo2014}.~Devices generating random numbers by exploiting the unpredictable nature of quantum processes are known as quantum random number generators (QRNGs) \cite{Schmidt1970,Miguel2016,Ma2016}.~Among all quantum physical systems, photons are possibly the most promising as they are easy to generate, manipulate and detect. Taking advantage of current photonics technology, QRNGs have been demonstrated based on the detection of single photons in different modes~\cite{Jennewein2000,Dynes2008,Wayne2010,Furst2010,Wahl2011,Applegate2015,Sanguinetti2014,Martin2015,Ma2016x}, quantum non-locality of entangled pairs of photons \cite{Pironio2010,Hugo2015}, phase noise of lasers \cite{Qi2010,Hong2010,Pan2015,Xu2012, Symul2011,Jofre2011,Carlos2015,Abellan2014,Yuan2014}, vacuum-seeded bistable processes \cite{Uchida2008,Marandi2012} and vacuum states \cite{Gabriel2010,Shi2016}. Yet, despite intense efforts to develop high-quality and high-speed QRNGs, more work is required for creating simple, cost-effective and practical devices.

In this paper, we propose and experimentally demonstrate a quantum random number generation scheme that is based on the creation of short laser pulses with quantum-random phases \cite{Lau1988}.
QRNGs based on such phase randomness have been demonstrated before: by interfering subsequent pulses in an unbalanced Mach-Zender interferometer (UMZI), the phase randomness was mapped onto easily-detectable intensity variations~\cite{Jofre2011,Carlos2015,Abellan2014,Yuan2014}. However, due to pulse emission-time jitter, the interference quality degrades significantly as the pulse length approaches the emission-time uncertainty, which limits the minimum pulse width and hence the maximum pulse rate~\cite{Yuan2014,Carlos2015}. In our scheme, the phase randomness of laser pulses is converted into intensity fluctuations by interfering them with another continuous wave laser featuring identical central frequency and polarization. 
The restriction of data acquisition to short time windows aligned -- possibly after pulse detection -- with the centres of the laser pulses effectively broadens and equalizes the spectra of the continuous wave laser and the pulsed laser, thereby ensuring high interference contrast even at high pulse repetition rates. 
Thus, our method not only inherently guarantees the temporal overlap needed for good interference, but can also create random numbers with narrower laser pulses and hence higher generation rates. Using only off-the-shelf opto-electronic and fiber-optic components at 1.5~$\mu$m wavelength, we perform a proof-of-principle experiment of the proposed scheme and extract high-quality quantum random numbers at a rate of 1.5 GHz. Moreover, we discuss ways to improve the performance, in particular the generation rate, of our scheme.

\section{Proposed scheme}
\label{Scheme}

\begin{figure}[ht]
\centering
\includegraphics[width=1\linewidth]{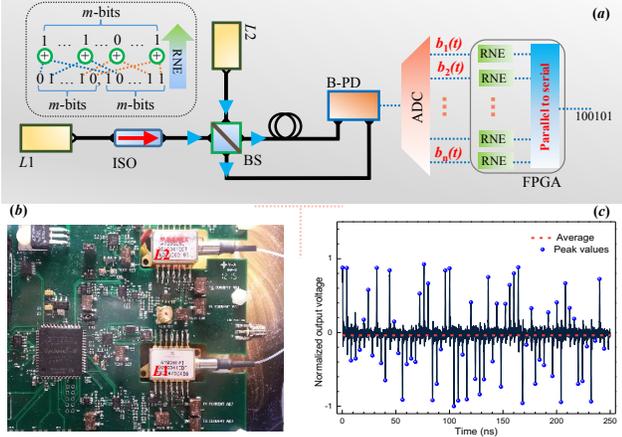}
\caption{(a) Schematic of our random number generator; (b) Picture of PCB board with gain-switched (pulsed) laser and (quasi)-continuous wave laser; (c) Typical signal from balanced-photo detector. \textit{L}1: gain-switched laser; \textit{L}2: (quasi)-continuous wave laser; ISO: optical isolator; BS: 50/50 beam splitter; B-PD: balanced-photo detector; ADC: analog to digital converter; RNE: randomness extractor; FPGA: field programmable gate array.}
\label{fig1}
\end{figure}
Figure~\ref{fig1}~(a) shows the idealized schematic of our random number generation.~A semiconductor laser,  \textit{L}1, is operated in gain-switched mode.~It is first biased far below threshold, i.e. around 0 mA, and then driven significantly above threshold using a short current pulse. This pulse samples and amplifies the vacuum fluctuation of the electromagnetic-field in the laser cavity, which results in the generation of laser pulses with quantum-random phases. Pulses from \textit{L}1 are then superposed with the output of a (quasi)-continuous wave laser, \textit{L}2, using a 50/50 beam splitter (BS). Note that an optical isolator (ISO) is used to avoid all light injecting into \textit{L}1, thereby preventing the generation of phase correlations between laser pulses \cite{Sun2015,Comandar2016}. 

The interfering pulses are detected by a balanced photo detector (B-PD). Ignoring detector noise, the differential voltage $\Delta V(t)$ output by the B-PD is
\begin{equation}
\Delta V(t) = 4\times \eta_{d}E_{1}(t)E_{2}(t)sin[\varphi_{1}(t)-\varphi_{2}(t)],
\label{eq1}
\end{equation}
where $\eta_{d}$ is the efficiency of the B-PD; $E_{1}(t)$, $E_{2}(t)$, $\varphi_{1}(t)$ and $\varphi_{2}(t)$ are the amplitudes and phases of the light fields from \textit{L}1 and \textit{L}2, respectively;~and $t = mT$, where $m$ is an integer and $T$ is the pulse period of~\textit{L}1. Since $\varphi_{1}(t)$ is random, electrical pulses of random amplitudes are obtained from B-PD.

To convert the pulses into raw bits, each pulse is input into an analog-to-digital converter (ADC) that divides the range of possible amplitudes into $2^{n}$ bins. (As we explain later, the maximum effective number of bins that can be used, $2^{n_{max}}$, is determined by the min-entropy of the signal from the B-PD~\cite{Jofre2011}.) With a voltage pulse from the B-PD as its input, the output of the ADC is specified by a vector with $n$ binary numbers as the elements and can be written as, 

\begin{equation}
O_{ADC} = (b_{1}, b_{2}, ...,b_{n})^{T},
\label{eq2}
\end{equation}

where $b_{x} = 0~or~1$, $x = 1, 2, ..., n$. Then we send the $n$ bits into a field programmable gate array (FPGA) that performs a randomness extraction procedure, resulting in true quantum-random bits. This procedure requires $n$ randomness extractors (RNEs).~Each RNE corresponds to one specific bit $b_{i}(t)$ per ADC output (see Fig.~\ref{fig1}~(a)).~Each RNE buffers $2m$ bits during $2m$ periods. All the bits buffered in the $n$ RNEs form a $n\times2m$ matrix in the FPGA which is,

\begin{equation}
B_{F0} = \begin{bmatrix}
    b_{1,1} & b_{1,2} & b_{1,3} & \dots  & b_{1,2m} \\
    b_{2,1} & b_{2,2} & b_{2,3} & \dots  & b_{2,2m} \\
    \vdots & \vdots & \vdots & \ddots & \vdots \\
    b_{n,1} & b_{n,2} & b_{n,3} & \dots  & b_{n,2m}
\end{bmatrix},
\label{eq3}
\end{equation}

Then the $n\times2m$ matrix divides them into two $n \times m$ matrices, such that,

\begin{equation}
B_{F1} = \begin{bmatrix}
    b_{1,1} & b_{1,2} & b_{1,3} & \dots  & b_{1,m} \\
    b_{2,1} & b_{2,2} & b_{2,3} & \dots  & b_{2,m} \\
    \vdots & \vdots & \vdots & \ddots & \vdots \\
    b_{n,1} & b_{n,2} & b_{n,3} & \dots  & b_{n,m}
\end{bmatrix},
\label{eq4}
\end{equation}

\begin{equation}
B_{F2} = \begin{bmatrix}
    b_{1,m+1} & b_{1,m+2}  & \dots  & b_{1,2m} \\
    b_{2,m+1} & b_{2,m+2} & \dots  & b_{2,2m} \\
    \vdots & \vdots  & \ddots & \vdots \\
    b_{n,m+1} & b_{n,m+2}  & \dots  & b_{n,2m}
\end{bmatrix}.
\label{eq40}
\end{equation}

The two $n \times m$ matrices are then XORed element wise, for e.g, $b_{1,1}$ with $b_{1,m+1}$, $b_{2,1}$ with $b_{2,m+1}$ and so on, as shown in the inset of~Fig.~\ref{fig1}~(a). This creates a $n \times m$ matrix as the output, as given in the Eq.~(\ref{eq5}).

\begin{equation}
\begin{split}
&   B_{F3}=  \\
&\begin{bmatrix}
b_{1,1}\oplus b_{1,m+1}& b_{1,2} \oplus b_{1,m+2}  & \dots  & b_{1,m} \oplus b_{1,2m}  \\
b_{2,1} \oplus b_{2,m+1} & b_{2,2} \oplus b_{2,m+2}  & \dots  & b_{2,m}\oplus b_{2,2m}  \\
\vdots & \vdots  & \ddots & \vdots \\
b_{n,1} \oplus b_{n,m+1} & b_{n,2} \oplus b_{n,m+2}    & \dots  & b_{n,m}\oplus b_{n,2m}
\end{bmatrix},
\label{eq5}
\end{split}
\end{equation}
where $\oplus$ represents the XOR operation.~The value of $m$ determines the separation between the two bits that are combined in the XOR gate. A larger $m$ means less correlation between bits.~Hence, with a proper value of $m$, the method presented here is equivalent to using two independent raw-bit sources, as demonstrated in Ref.~\cite{Uchida2008}. Given that our randomness extraction procedure is arguably information-theoretic secure, the quality of the randomness of the extracted bits is tested using the standard NIST test suite as shown in section~\ref{Results} alongside a  measurement of the auto-correlation of the bits before and after extraction, shown in Fig~\ref{auto-correlation}. Finally, after parallel-to-serial conversion, the bits from all RNEs form a string of ready-to-use random bits. Thus we can achieve an average generation rate of random numbers of $nR/2$, where $R=1/T$ is the repetition rate of the pulsed laser \textit{L}1.~We note that, compared with randomness extraction using a cryptographic hash function~\cite{Nisan1999}, the employed RNE method in our scheme imposes less performance on the FPGA and is much easier to implement in real time. However, it may result in losing more random bits than necessary to obtain a final quantum-random bit string. 

\section{Proof-of-principle demonstration}
\label{Experiment}
Figure 1 (b) shows a picture of the laser drivers and lasers \textit{L}1 and \textit{L}2 used in our experimental demonstration of the proposed scheme.~The central wavelengths of both lasers are at 1540 nm -- they are matched and stabilized by controlling the temperature of lasers within 0.01~\degree C. The gain-switched laser is driven by a sequence of current pulses that are generated from a radio-frequency transistor switched on/off by an FPGA signal. The width of the current pulse is $\sim$200 ps, and the repetition rate is 250 MHz.~After interference with the output from the (quasi)-continuous wave laser \textit{L}2 in a polarization maintaining 50/50 BS (used to match the polarization mode, thus maximize the visibility of interference), the optical signals are then detected by a commercial B-PD (Thorlabs, PDB480C). It is worth noting that the balanced detection scheme removes all common-mode noise, which results in the improvement of the signal-to-noise ratio of the detection signal.~Figure~\ref{fig1}~(c) shows typical signals from B-PD, i.e.~$\Delta V(t)$ given in Eq.~(\ref{eq1}). The dashed line is the average of the detected signal. Please note that, in our proof-of-principle demonstration, the ADC, RNEs and parallel-to-serial conversion described above have not been implemented using an FPGA. Instead, we used a computer to process analog signals from B-PD that have previously been sampled by a fast oscilloscope (Lecroy, 8600A). Hence, while we demonstrate a proof-of-principle of the proposed scheme, the random numbers are not yet generated in real time.
	
\section{Results}
\label{Results}
\begin{figure}[ht]
\centering
\includegraphics[width=1\linewidth]{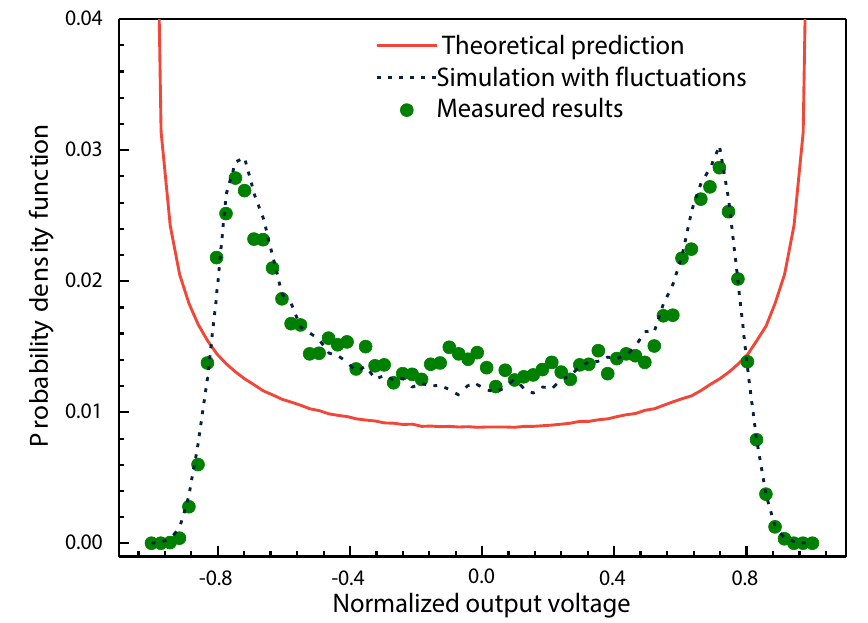}
\caption{Probability density function of the normalized analog signals, $\Delta V(t)$}.
\label{fig2}
\end{figure}
As shown in Eq.~(\ref{eq1}), the phase uncertainty of the emitted laser pulses affects $\Delta V(t)$ through the interference and balanced photo-detection. Figure \ref{fig2} shows the probability density function (PDF) of the normalized $\Delta V(t)$, sampled at a pulse center. The dots represent the experimental results. The solid red line is the theoretical prediction of the corresponding PDF which is,

\begin{equation}
p(x)= 1/(\pi\sqrt{1-x^{2}}),
\label{eq6}
\end{equation}
where $x$ is the normalized analog output of the B-PD, and the phase distribution is assumed to be uniform. We attribute the deviation of our experimental results from the theoretical prediction to additional amplitude fluctuations in the detection signal that stem from classical sources, such as peak power fluctuations of laser pulses, limited bandwidth of the B-PD, the finite sampling rate and the noise of the oscilloscope. We estimate the extent of these amplitude fluctuations by inputting the laser pulses from \textit{L}1 into one of the photo-detectors of the B-PD and analyzing its output using the same oscilloscope. Ideally, without the above-mentioned fluctuations, we would expect a constant output from that detector. However, we found an electrical signal whose amplitude follows a Gaussian distribution with standard deviation of $\leq$~5\% compared to the full range of the observed electrical signal. We simulate the effect of these classical fluctuations by adding them to the predicted values for the ideal case using a Monte-Carlo method.~The dashed line in Fig.~\ref{fig2} shows the good agreement of the result with the measured data.~This allows us not only to verify that the amplitude of each pulse is indeed random (but not fully quantum-random), but also suggests ways to improve the quality of the random numbers, such as using a B-PD and ADC with larger bandwidth. 

One of the main advantages of this random number generation scheme is that more than one random bit can be obtained per detection. The total range of the measured signal can be divided into $2^n$ bins, and each signal represented by $n$ bits. The maximum number of bits, $n_{max}$, that can be extracted is determined by the min-entropy of the analog signal from B-PD,  
\begin{equation}
\label{eq2}
H_{min} = -{log}_2 (p_{max})
\end{equation}
where $p_{max}$ is the maximum probability for the detection amplitude to belong into any of the $2^n$ bins. By increasing the number of bins, we find that $H_{min}$ saturates at 12.8 for $n \geq$ 13, indicating that $p_{max} = 2^{-12.8}$ and $n_{max}=$~12 raw random bits can be extracted from each pulse \cite{Jofre2011}. 

However these pulses contain entropy from both quantum and classical sources. To estimate contribution from quantum noise, the $p_{max}$ of the quantum noise, quantum min-entropy, is $2^{-6.49}$, which was estimated following the procedure in \cite{Abellan2014}. We note that for this calculation, the classical noise is assumed to be independent of quantum noise. The metrological approach to quantify the randomness will be applied for future demonstrations~\cite{Carlos2015}.We find an $H_{min}$ of 6.49, indicating that $n_{max}=$~6 quantum random bits can be extracted from each pulse. 
To improve the quality of randomness, we employ the randomness extraction procedure described in section~\ref{Scheme}, which reduces the information per laser pulse from 12 to 6 bits. 
Although this satisfies the approved cryptographic conditioning components set forth by NIST \cite{NIST-RNG}, we also used a more standard randomness extraction procedure, i.e. the Toeplitz hashing matrix, to validate the tests we perform. This matrix was set up to similarly reduce the information per laser pulse from 12 to 6 bits. The following procedure was performed on both sets of data.
Therefore, with a clock rate of 250~MHz, 12-bit binning and the randomness extraction, random bits are obtained at 1.5 GHz, which is half of the maximum of 3.0 GHz = $12 \times 250$ MHz.

\begin{figure}[ht]
\centering
{\includegraphics[width=1\linewidth]{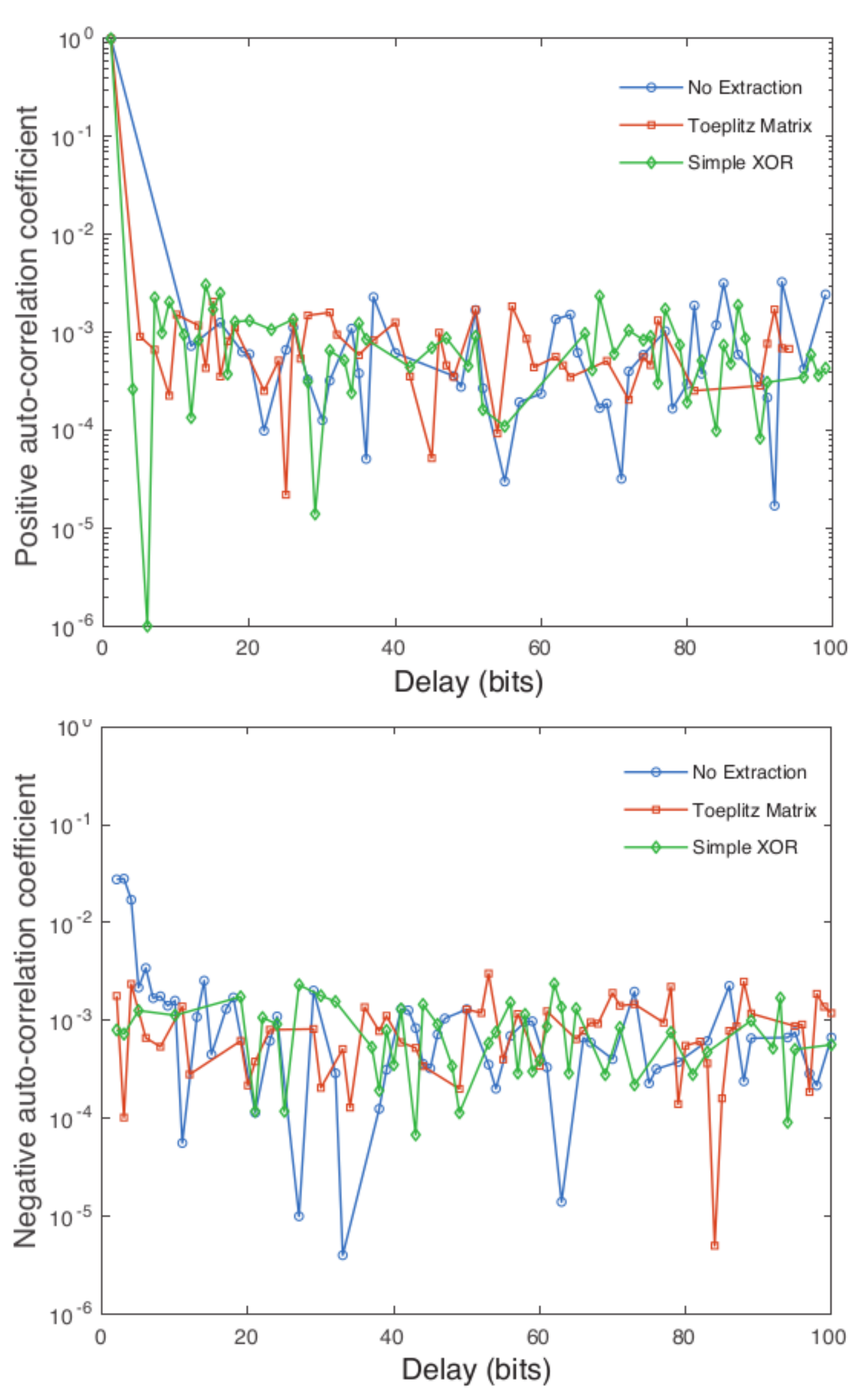}}
\caption{Auto-correlation results for the random bits before and after Hashing- or XORed- extraction.}
\label{auto-correlation}
\end{figure}

\begin{figure}[ht]
\centering
{\includegraphics[width=1\linewidth]{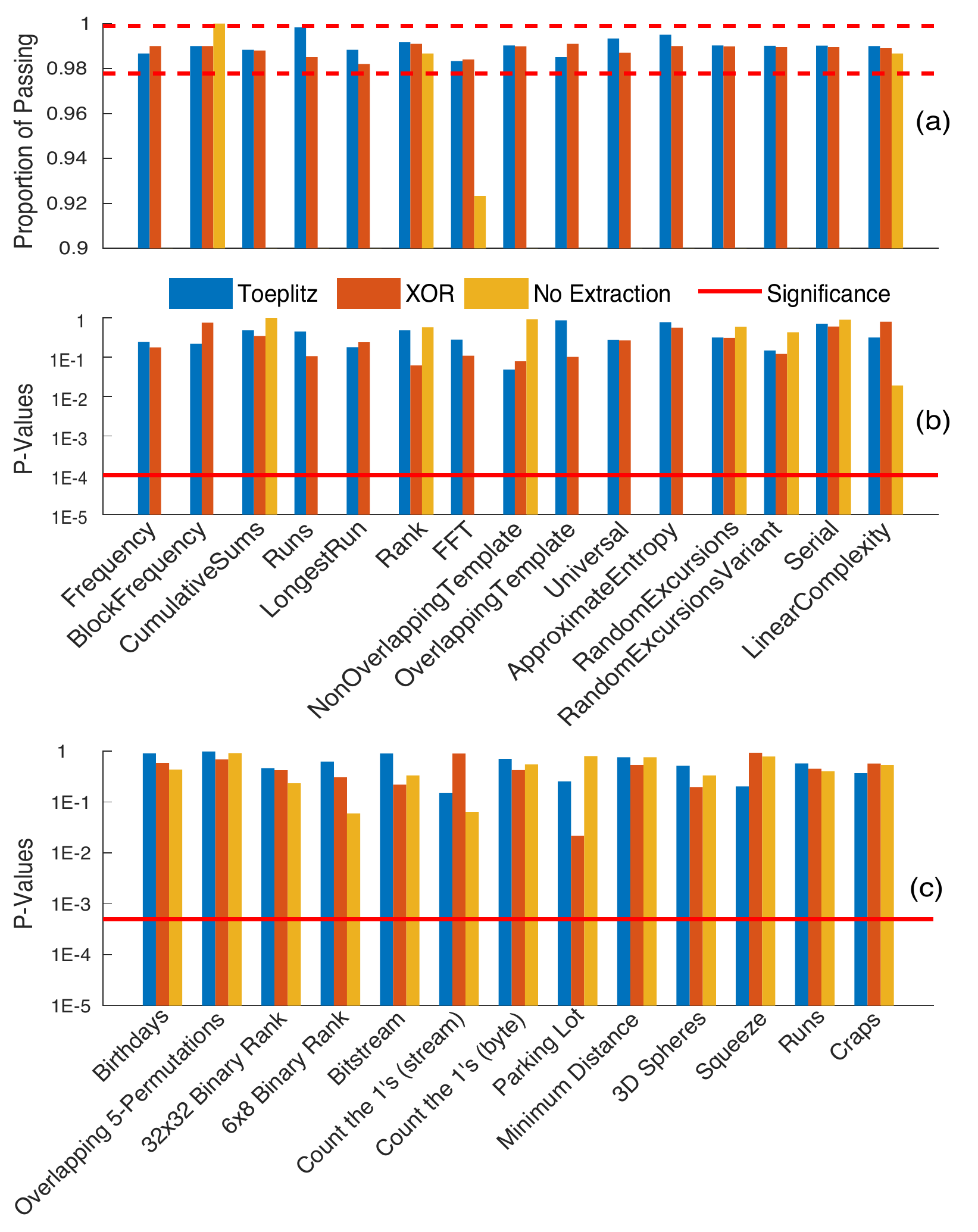}}
\caption{Results of the NIST and Dieharder tests applied to ~1.25 Gbits of random bits. (a) The proportion of passes of each test in the NIST suite for 600 1-Mb-long samples. All tests are passed with a proportion value greater than 0.9778 and less than 1;~(b) the P-values of each individual (NIST) test, obtained from the distribution of P-values of each of the 600 trials. For the tests, which produce multiple P-values and proportions, the worst cases are given.~(c) the P-values of a select number of tests from the Dieharder test suite. All tests are passed with 600 1-Mb-long samples and at a significance level of 0.0001 for the NIST tests and 0.0005 for the DIEHARDER tests. }
\label{fig3}
\end{figure}

To assess the quality of the final random bits obtained from our setup, we first create a ~1.25 Gbit-long random file by saving measurement results from the oscilloscope and processing them in a computer. We measure auto-correlation, see~\cite{Xu2012} for the formula used, of the processed random bits before and after randomness extraction with the Toeplitz hashing matrix or XOR operation (with $m = 7$ explained later), and the results are shown in Fig.~\ref{auto-correlation}. As can be seen, both extraction procedures bring the correlation of the first few bits down to the baseline level. We also subject the random bits to 2 statistical suites; The NIST STS (Statistical Test Suite) which is a battery of fifteen tests used to analyze the statistical properties of random numbers~\cite{Rukhin2008}, and the DIEHARDER~\cite{Dieharder} battery of tests which has the same goals but is developed independantly of NIST. By monitoring the results of the NIST test as a function of $m$ (i.e. the length of the buffer in the RNEs), we find that with $m = 7$, the obtained random file passes all the tests. It is worth emphasizing that these test results do not mean that our source is truly random, they can only assess the properties of the source~\cite{NIST-RNG}.

For the NIST test, the significance level ($\alpha$) is set at 0.01 as suggested by the test suite~\cite{Rukhin2008}, implying that one out of one hundred tests is expected to fail even if the random numbers being tested are generated by a fair random generator. Each of the fifteen tests is considered a pass if the proportion of success versus fail is within a range given by $\hat{p}\pm3\sqrt{\hat{p}(1-\hat{p})/{N}}$, where $N$ is the number of times an individual test runs ($N$= 600 in our case), and $\hat{p}=1-\alpha$. This results in the proportion value greater than 0.9778 and less than 1, a range that is indicated by the two dashed lines in Fig. \ref{fig3}~(a). 
 Next, a P-value is obtained for each test from the distribution of P-values over 600 trials. It is considered a pass if this P-value is above the suggested significance level of 0.0001~\cite{Abellan2014}. As shown in Fig.~\ref{fig3}, the random numbers from either our XOR method or the standard randomness extraction method pass all the NIST tests.~Our result shows that both our scheme to extract the randomness by sampling vacuum fluctuations and the XOR method to extract the quantum random bits are feasible, thus paves a practical avenue to obtain quantum random bits with good quality.
 
\section{Conclusion}
\label{Conclusion}

We introduced and reported a proof-of-principle demonstration of a new scheme for creating high-quality quantum-random bits based on a gain-switched and a (quasi)-continuous wave laser. The generation rate, currently 1.5 Gbps, can be further increased by operating the gain-switched laser with higher repetition rate. While this rate is fundamentally limited due to the need for laser cavity depletion in-between subsequent pulses, rates of several GHz for gain-switched laser are feasible \cite{Yuan2014,Abellan2014}. Combined with the possibility to create more than 10 random bits per laser pulse, we therefore predict that our scheme can deliver high-quality quantum random numbers at rates of many tens of GHz. We note that, while the present work was being finalized, a related experimental demonstration using a photonics chip has been reported ~\cite{Abellan2016, comment}. 

\section*{Acknowledgments}

This work was funded through Alberta Innovates Technology Futures (AITF), and the National Science and Engineering Research Council of Canada (NSERC). WT furthermore acknowledges funding as a Senior Fellow of the Canadian Institute for Advanced Research (CIFAR). QZ also acknowledges funding from the National Key R\&D Program of China (No. 2018YFA0307400), and National Nature Science Foundation of China (No. 61775025 and No. 61405030).
The authors thank Vladimir Kiselyov for technical support, and Daniel Oblak and Carlos Abell\'an for useful discussions.

\end{document}